\documentclass[prl,twocolumn,showpacs,superscriptaddress,amsfonts,amsmath,
floatfix]{revtex4}
\usepackage{graphicx}
\usepackage{psfrag}%To include LaTeX subscripts on figures
\newcommand{\Journal}[4]{#1 {\bf #2}, #3 (#4)}
\newcommand{\PR}{Phys. Rev.}
\newcommand{\PRL}{Phys. Rev. Lett.}
\newcommand{\PRA}{Phys. Rev. A}

\newcommand{\JMP}{J. Math. Phys.}

\newcommand{\Science}{Science}

\newcommand{\PLA}{Phys. Lett. A}
\begin{document}
\title {Bosonization, Pairing, and Superconductivity of the Fermionic Tonks-Girardeau Gas}
\author{M. D. Girardeau}
\email{girardeau@optics.arizona.edu}
\affiliation{College of Optical
Sciences, University of Arizona,
Tucson, AZ 85721, USA}
\author{A. Minguzzi}
\email{anna.minguzzi@grenoble.cnrs.fr}
\affiliation{Laboratoire de Physique et Mod\'elisation des Mileux Condens\'es, 
C.N.R.S., B.P. 166, 38042 Grenoble, France}
\affiliation{Laboratoire de Physique Th\'{e}orique et Mod\`{e}les
Statistiques, Universit\'{e} Paris-Sud, B\^{a}t. 100, F-91405 Orsay, France}
%\date{\today}
%
\begin{abstract}
We determine some exact static and time-dependent properties of the fermionic
Tonks-Girardeau (FTG) gas, a spin-aligned one-dimensional Fermi gas  
with infinitely strongly attractive zero-range odd-wave interactions. We show 
that the two-particle reduced density matrix exhibits maximal superconductive
off-diagonal long-range order, and on a ring an FTG gas with an even number 
of atoms has a highly degenerate ground state with quantization of 
Coriolis rotational flux and high sensitivity to rotation and to external
fields and accelerations.
For a gas initially under harmonic confinement we show that during an expansion
the momentum distribution undergoes a ``dynamical bosonization'', approaching
that of an ideal Bose gas without violating the Pauli exclusion principle.
\end{abstract}
\maketitle
If an ultracold atomic vapor is confined in a de Broglie wave guide with 
transverse trapping so tight and temperature so low that the transverse
vibrational excitation quantum $\hbar\omega$ is larger than available
longitudinal zero point and thermal energies, the effective dynamics becomes   
one-dimensional (1D) \cite{Ols98,PetShlWal00}, a regime currently under 
intense experimental
investigation \cite{Tol04Mor04,Par04Kin04}. Confinement-induced 1D 
Feshbach resonances (CIRs) reachable by tuning the 1D 
coupling constant via 3D Feshbach scattering resonances 
occur for both Bose gases \cite{Ols98} and spin-aligned Fermi gases 
\cite{GraBlu04}. Near a CIR the 1D interaction is very 
strong, leading to strong short-range correlations, breakdown of 
effective-field theories, and emergence of highly-correlated $N$-body
ground states. In the bosonic case with very strong repulsion (1D hard-core
Bose gas with coupling constant $g_{1D}^{B}\to +\infty$, the 
Tonks-Girardeau (TG) gas), the exact 
$N$-body ground state was determined some 45 years ago by a 
Fermi-Bose (FB) mapping to an ideal Fermi gas \cite{Gir60}, leading to 
``fermionization'' of many properties of this Bose system, as recently 
confirmed experimentally \cite{Par04Kin04}. 
The ``fermionic TG'' (FTG) gas \cite{GirOls03GirNguOls04}, a spin-aligned Fermi gas 
with very strong \emph{attractive} 1D odd-wave interactions, can be realized by
3D Feshbach resonance mediated tuning to the attractive side of the CIR with 1D
coupling constant $g_{1D}^{F}\to -\infty$. It has been pointed out 
\cite{GraBlu04,GirOls03GirNguOls04} that the generalized FB mapping
\cite{CheShi98,GraBlu04,GirOls03GirNguOls04} can be exploited in the opposite
direction to map this system to the trapped \emph{ideal Bose} gas, 
leading to determination of the exact $N$-body ground state and 
``bosonization'' of many properties of this Fermi system. We recently 
examined the equilibrium one-body density matrix and 
exact dynamics following sudden turnoff of the interactions by detuning from 
the CIR \cite{GirWri05}. Here we determine some other exact properties of  
the untrapped, ring-trapped, and harmonically trapped fermionic 
TG gas, the most striking of which are pairing, superconductive off-diagonal 
long-range order (ODLRO) of the two-body density matrix, a highly degenerate
ground state of an even number of atoms on a ring with quantization of 
Coriolis rotational flux and high sensitivity to rotation and to external
fields and accelerations, and a ``dynamical
bosonization'' of the momentum distribution following sudden relaxation
of the trap frequency. 

{\it Untrapped FTG gas:} The Hamiltonian is
$\hat{H}=\sum_{j=1}^{N}
\left[-\frac{\hbar{^2}}{2m}\frac{\partial^2}{\partial x_{j}^{2}}\right]
+\sum_{1\le j<\ell\le N}v_\text{int}^{\text{F}}(x_{j}-x_{\ell})$
where $v_\text{int}^{\text{F}}$ is the two-body interaction. Since the spatial
wave function is antisymmetric due to spin polarization, there is no 
s-wave interaction, but it has been shown
\cite{GraBlu04,GirOls03GirNguOls04} that a strong, attractive, 
short-range odd-wave interaction (1D analog of 3D p-wave interactions) 
occurs near the CIR. This can be modeled
by a narrow and deep square well of depth $V_0$ and width $2x_0$. The contact
condition at the edges of the well is \cite{GirOls03GirNguOls04}
$\psi_{F}(x_{j\ell}=x_{0})=-\psi_{F}(x_{j\ell}=-x_{0})
= -a_{1D}^{F}\psi_{F}^{'}(x_{j\ell}=\pm x_{0})$
where $a_{1D}^{F}$ is the 1D scattering length and the prime denotes 
differentiation. Consider first the relative wave function $\psi_{F}(x)$
in the case $N=2$. The FTG limit is 
$a_{1D}^{F}\to -\infty$, a zero-energy scattering resonance.
The exterior solution is $\psi_{F}(x)=\text{sgn}(x)=\pm 1$ 
($+1$ for $x>0$ and $-1$ for $x<0$) and the
interior solution fitting smoothly onto this is $\sin(\kappa x)$ with
$\kappa=\sqrt{mV_{0}/\hbar^2}=\pi/2x_0$. In the zero-range limit
$x_{0}\to 0+$ the well area $2x_{0}V_{0}=(\pi\hbar)^{2}/2mx_{0}\to\infty$,
stronger than a negative delta function. In this limit the 
wave function is discontinuous at contact $x_{0}=0\pm$, allowing an infinitely 
strong zero-range interaction in spite of the
antisymmetry of $\psi_F$ \cite{CheShi98}. This generalizes immediately to
arbitrary $N$: the exact FTG gas ground state is 
\begin{equation}
\label{eq:psiF}
\psi_{F}(x_{1},\cdots,x_{N})=
A(x_{1},\cdots,x_{N})\prod_{j=1}^{N}\phi_{0}(x_{j})
\end{equation} 
 with
$A(x_{1},\cdots,x_{N})=\prod_{1\le j<\ell\le
N}\text{sgn}(x_{\ell}-x_{j})$ the ``unit antisymmetric function''
employed in the original discovery of fermionization \cite{Gir60} and
$\phi_0=1/\sqrt{L}$ the ideal Bose gas ground orbital, $L$ being the
periodicity length. Its energy
is zero \cite{Note1} and it satisfies periodic boundary conditions for odd 
$N$ and antiperiodic boundary conditions for even $N$ \cite{commentbc}. 

The exact single-particle density matrix 
$\rho_{1}(x,x')=N\int\psi_{F}(x,x_{2},\cdots,x_{N})
\psi_{F}^{*}(x',x_{2},\cdots,x_{N})dx_{2}\cdots dx_N$ is 
\cite{BenErkGra04,GirWri05} 
$\rho_1(x,x')=N\phi_{0}(x)\phi_{0}^{*}(x')[F(x,x')]^{N-1}$ with
$F(x,x')=\int_{-L/2}^{L/2}\text{sgn}(x-y)\text{sgn}(x'-y)|\phi_{0}(y)|^{2}dy
=1-2|x-x'|/L$. In the thermodynamic limit
$N\to\infty$, $L\to\infty$, $N/L=n$ this gives an exponential decay 
\cite{BenErkGra04}: $\rho_1(x,x')=ne^{-2n|x-x'|}$. Its Fourier transform 
$n_k$, normalized to $\sum_{k}n_{k}=N$ (allowed momenta $\nu 2\pi/L$ with
$\nu=0,\pm 1,\pm 2,\cdots$), is the momentum distribution function
$n_{k}=[1+(k/2n)^2]^{-1}$. It satisfies the exclusion principle limitation 
$n_{k}\le 1$, but nevertheless, for $n\to 0$ the 
continuous momentum density $n(k)=(L/2\pi)n_k$ reduces to $N$
times a representation of the Dirac delta function, simulating the 
ideal \emph{Bose} gas distribution: 
$n(k)\longrightarrow_{_{\hspace{-.6cm}n\to 0}}N\delta(k)$ 
\cite{BenErkGra04}. 

The two-particle density matrix $\rho_{2}(x_{1},x_{2};x_{1}',x_{2}')
= N(N-1)\int \psi_{F}(x_{1},..,x_{N})
\psi_{F}^{*}(x_{1}',x_{2}',x_{3},..,x_{N}) dx_{3}...dx_{N}$ 
also has a simple closed form: 
\begin{eqnarray}
\label{eq:rho2}
& &\rho_{2}(x_{1},x_{2};x_{1}',x_{2}')=
N(N-1)\text{sgn}(x_{1}-x_{2})\phi_{0}(x_{1})\phi_{0}(x_{2})\nonumber\\
&\times&\text{sgn}(x_{1}'-x_{2}')
\phi_{0}^{*}(x_{1}')\phi_{0}^{*}(x_{2}')
[G(x_{1},x_{2};x_{1}',x_{2}')]^{N-2}
\end{eqnarray}
where $[G(x_{1},x_{2};x_{1}',x_{2}')]^{N-2}
=[\int_{-L/2}^{L/2}\text{sgn}(x_{1}-x)\text{sgn}(x_{2}-x)
\text{sgn}(x_{1}'-x)\text{sgn}(x_{2}'-x)|\phi_{0}|^{2}(x)dx]^{N-2}
=e^{2n(y_{1}-y_{2}+y_{3}-y_{4})}$ in the thermodynamic limit and 
$y_{1}\le y_{2}\le y_{3}\le y_{4}$ are the arguments 
$(x_{1},x_{2};x_{1}',x_{2}')$ in ascending order. 
$\rho_2$ is of order $n^2$ in the following cases: (a) 
$|x_{1}-x_{1}'|\le \text{O}(1/n)$,\ $|x_{2}-x_{2}'|\le \text{O}(1/n)$; (b) 
$|x_{1}-x_{2}'|\le \text{O}(1/n)$,\ $|x_{2}-x_{1}'|\le \text{O}(1/n)$; (c) 
$|x_{1}-x_{2}|\le \text{O}(1/n)$,\ $|x_{1}'-x_{2}'|\le \text{O}(1/n)$. 
These are just Yang's criteria \cite{Yan62} for superconductive ODLRO of 
$\rho_2$ in the absence of ODLRO of $\rho_1$. In case (c) $\rho_2$ remains
of order $n^2$ for arbitrarily large separation of the centers of
mass $X=(x_{1}+x_{2})/2$ and $X'=(x_{1}'+x_{2}')/2$, the hallmark
of ODLRO. On the other hand, in cases (a) and (b) $\rho_2$
decays exponentially with $|X-X'|$. In the thermodynamic limit only 
configurations (c) contribute to the largest eigenvalue of $\rho_2$,
and $\rho_2$ separates apart from negligible contributions (a) and (b)
\cite{Note2}:
\begin{eqnarray}
\rho_{2}(x_{1},x_{2};x_{1}',x_{2}')
=n^{2}\text{sgn}(x_{1}-x_{2})e^{-2n|x_{1}-x_{2}|}\nonumber\\
\times\ \text{sgn}(x_{1}'-x_{2}')e^{-2n|x_{1}'-x_{2}'|}
+\text{terms negligible for }\lambda_{1}.
\end{eqnarray}
By Yang's argument \cite{Yan62} the largest eigenvalue is $\lambda_{1}=N$,
and this is confirmed by comparison with the $\lambda_1$ contribution 
$\lambda_{1}u_{1}(x_{1},x_{2})u_{1}(x_{1}',x_{2}')$ to the spectral 
representation of $\rho_{2}$, implying that the corresponding 
eigenfunction is $u_{1}(x_{1},x_{2})
=\mathcal{C}\ \text{sgn}(x_{1}-x_{2})e^{-2n|x_{1}-x_{2}|}$ with \cite{Note3} 
$\mathcal{C}=\sqrt{n/L}$, confirming the value 
$\lambda_{1}=n^{2}/{\mathcal{C}}^{2}=N$.  
The range $1/2n$ of $u_1$ is in the region of onset 
of a BEC-BCS crossover between tightly bound bosons and loosely bound Cooper
pairs. There is an upper bound \cite{Yan62}
$\lambda_{1}\le N$ on the largest eigenvalue, so 
\emph{the untrapped FTG gas is maximally superconductive} in the
sense of Yang's ODLRO criterion. 

{\it FTG gas on a ring:}  If the FTG gas is trapped on a 
circular loop of radius $R$, with particle
coordinates $x_j$ measured around the circumference $L=2\pi R$, the
FTG gas must satisfy periodic boundary conditions for both odd and even
$N$ because of single-valuedness of its  wave function. Since the
mapping function  $A(x_{1},\cdots,x_{N})=\prod_{1\le j<\ell\le
N}\text{sgn}(x_{\ell}-x_{j})$ is periodic (antiperiodic) for odd (even) 
$N$ as a result of its definition, it follows that the mapped
ideal Bose gas used to solve the FTG problem must satisfy periodic
(antiperiodic) boundary conditions 
for odd (even) $N$. The ground state of a FTG
gas on a ring is then different depending on the particle number parity. 
For odd $N$ the FTG ground state in Eq.~(\ref{eq:psiF}) is built from
 the zero-momentum orbital $\phi_0=1/\sqrt{L}$ and corresponds to mapping the
FTG gas onto the ideal Bose gas ground state, the usual complete 
Bose-Einstein condensate (BEC), and is nondegenerate.
On the other hand, for even $N$, which we henceforth assume, antiperiodicity
requires that the only  plane-wave orbitals allowed are 
 $e^{ikx_j}/\sqrt{L}$ with $k=\pm\pi/L,\pm 3\pi/L,\cdots$. The
ground state of this fictitious ideal Bose gas, and hence that of the    
mapped FTG gas, is then $(N+1)$-fold degenerate, with energy eigenvalue
$N(\hbar^2/2m)(\pi/L)^2$. These degenerate ground states
are fragmented BECs with $wN$ atoms in the orbital $e^{i\pi x_j/L}$
and $(1-w)N$ in $e^{-i\pi x_j/L}$ with $0\le w\le 1$, and are conveniently 
labelled by a quantum number 
$\ell_z=(w-\frac{1}{2})N=0,\pm 1,\pm 2,\cdots,\pm \frac{N}{2}$ 
related to the eigenvalue
$P$ of circumferencial linear momentum and that $L_z$ of angular momentum
$z$-component by $P=\ell_z \hbar/R$ and $L_z=\ell_z\hbar$. The angular
momentum per particle is half-integral due to
antiperiodicity of the orbitals, and the degenerate ground states 
are in one-one correspondence with the eigenstates of \emph{spin}
angular momentum $z$-component of $N$ spin-$1/2$ fermions.

The ground state degeneracy makes the FTG gas on a ring a good candidate
for detecting small external fields and linear accelerations. Suppose that
there is a potential gradient parallel to a diameter of the 
ring, or an acceleration leading
to a gradient in the inertial potential arising from Einstein's
principle of equivalence, with the circumferential minimum
of this potential occurring at a point $x_0$. Then the degeneracy 
is lifted and to lowest order in degenerate perturbation theory 
all $N$ atoms occupy 
the orbital $\phi_0(x)=\sqrt{2/L}\cos[\pi(x-x_0)/L]$, leading to an  
observable asymmetric density profile $n(x)=2n\cos^2[\pi(x-x_0)/L]$. 

\begin{figure}
%[htbp]
  \centering
  \psfrag{N hbar^2/2mR^2}{$E$ $[N\hbar^2/2mR^2]$}  
  \psfrag{Phi/Phi0}{$\Phi/\Phi_0$} 
\psfrag{lz=-5N/2}{$\frac{-5N\hbar}{2}$} 
\psfrag{lz=-3N/2}{$\frac{-3N\hbar}{2}$} 
\psfrag{lz=-N/2}{$\frac{-N\hbar}{2}$} 
\psfrag{lz=N/2}{$\frac{N\hbar}{2}$} 
\psfrag{lz=3N/2}{$\frac{3N\hbar}{2}$}
\psfrag{lz=5N/2}{$\frac{5N\hbar}{2}$}
\includegraphics[width=7.5cm,angle=0]{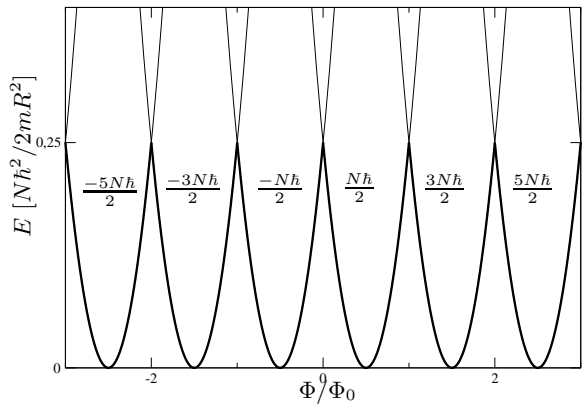} 
  \caption{Dependence of energies $E$ on rotational flux $\Phi$.
Heavy line: Ground state energy $E_0(\Phi)$. Lighter lines: Lowest energy
for each value of total angular momentum.}
  \label{fig1}
\end{figure}

Due to its quantum coherence the FTG gas is also a good candidate 
for a sensitive rotation detector. Suppose that the 
ring trap is rotating with angular velocity  $\vec{\omega}$
perpendicular to the plane of the ring. In the rotating 
coordinate system each atom sees an effective Coriolis force
$\vec{F}_{\text{Cor}}=2m\vec{v}\mathbf{\times}\vec{\omega}$.
Comparing this with the usual magnetic force 
$\vec{F}_{\text{mag}}=(e/c)\vec{v}\mathbf{\times}\vec{B}$, one sees that
the kinetic energy operators in the Hamiltonian in the rotating system
are $[\hat{p}_j-\frac{h}{L}\frac{\Phi}{\Phi_0}]^2/2m$ where 
$\hat{p}_j=(\hbar/i)\partial/\partial x_j$, $\Phi=\pi R^2\omega$ is the
Coriolis flux through the loop, and $\Phi_0=h/2m$ is the Coriolis flux
quantum. The energy of each state $|\ell_z\rangle$ then becomes
$E=E_0(\Phi=0)+\frac{N\hbar^2}{2mR^2}[(\frac{\Phi}{\Phi_0})^2
-2\ell_z\frac{\Phi}{\Phi_0}]$ which is minimized when 
$\ell_z=\frac{N}{2}$ if $\Phi>0$ and $\ell_z=-\frac{N}{2}$ if $\Phi<0$,
i.e., even a very small angular velocity leads to a nondegenerate ground state
with all $N$ atoms at either $k=\pi/L$ or $k=-\pi/L$. Generalizing to states 
differing from the $\Phi=0$ ground states by displacement in $k$-space
by integral multiples of $2\pi/L$ one obtains the $\Phi$-dependent
ground state energy $E_0(\Phi)$ shown by the heavy line in Fig. \ref{fig1},
in which the lighter lines show the lowest energies for 
$\ell_z=\pm\frac{N}{2},\pm\frac{3N}{2},\cdots.$ The ground
state energy is a periodic function of $\Phi$ with period 
$\Phi_0$ in accord with a general theorem \cite{Yan62}, but unlike the usual 
situation for a superconductor, (a) there is no smaller period $\Phi_0/2$, 
and (b) for even $N$, $\Phi_0=0$ is a relative maximum  of $E_0$ rather than a
minimum (as is the case of odd $N$),
the first minima occuring at $\Phi=\pm\Phi_0/2$.
%**********************************************************************
%**********************************************************************
The barrier heights of the energy landscape in Fig.\ref{fig1} vanish 
like $1/N$ for $N\to\infty$, so flux quantization
will not be observable for a macroscopic ring. However, it may be observable
for mesoscopic rings using BEC-on-a-chip technology. For example, assuming
a ring radius $R=$5 $\mu$m, one finds that for $^6\mbox{Li}$, 
$\Delta E>k_B T$ for $T<50$ nK.  

{\it Flow properties on a nonrotating ring:} 
According to the FB
mapping the excitation spectrum of the FTG gas is the same as that of
an ideal Bose gas, and hence it is sufficient to analyze the latter.
Since the excitation energy of the
ideal Bose gas is quadratic in the excitation momentum $\hbar q$,
the FTG gas  does not satisfy the Landau-Bogoliubov criterion for 
superfluidity. We investigate here the possibility
of  flow metastability associated with barriers in the
excitation energy landscape as a function of the transferred momentum. 
It was shown by F. Bloch
\cite{Blo73} that for the usual ideal Bose gas, which corresponds to the 
case of odd $N$ in our treatment, no such barriers exist. In the case of even $N$ both the ground state and the excitation branches are $(N+1)$-fold
degenerate, but it is sufficient here to consider
the $\ell_z=0$ ground state and the excitations arising from it  by promoting 
atoms to higher $k$-values. Generalizing Bloch's analysis, 
we note that for $0<\nu\le N/2$ the lowest branch corresponds to excitation
of $\nu$ atoms from $k=-\pi/L$ to $k=3\pi/L$, yielding a state with 
angular momentum $z$-component $\ell_z\hbar$ with $\ell_z=2\nu$, and 
with excitation energy $\epsilon(\ell_z)=\ell_z\hbar^2/mR^2$. At $\nu=N/2$
one has reached a state differing from the ground state by translation
of all atoms by an amount $2\pi/L$ in $k$-space, and one can repeat
this process, promoting atoms from $k=\pi/L$ to $5\pi/L$,
yielding another straight-line segment connecting the points $\ell_z=N$
and $\ell_z=2N$ on a parabolic curve $(\ell_z\hbar)^2/2NmR^2$, etc.
Together with symmetry $\epsilon(\ell_z)=\epsilon(-\ell_z)$ this yields an 
excitation energy curve composed of straight-line segments as in the dashed curve of Bloch's Fig. 2 \cite{Blo73} with the notation
$P=\ell_z\hbar/R$. Hence for both odd and even $N$
there are no energy barriers, and 
the FTG gas on a nonrotating ring does not exhibit flow metastability.

\begin{figure}
%[htbp]
  \centering
  \psfrag{p/p_ho}{$k/k_{osc}$} 
  \psfrag{t=0}{$t=0$ $\longrightarrow$} 
\psfrag{t=0}{$t=0$ $\longrightarrow$} 
\psfrag{t=5}{$t=5$ $\longrightarrow$} 
\psfrag{t=10}{$t=10$ $\longrightarrow$} 
\psfrag{t=20}{$t=20$ $\longrightarrow$} 
\psfrag{t=30}{$t=30$ $\longrightarrow$} 
\psfrag{t=40}{$t=40$ $\longrightarrow$}
\psfrag{ideal Bose gas}{ideal Bose gas $\longrightarrow$} 
\psfrag{n(p,t) p_ho}{$n(k,t)\, k_{osc}$}
\includegraphics[width=7.5cm,angle=0]{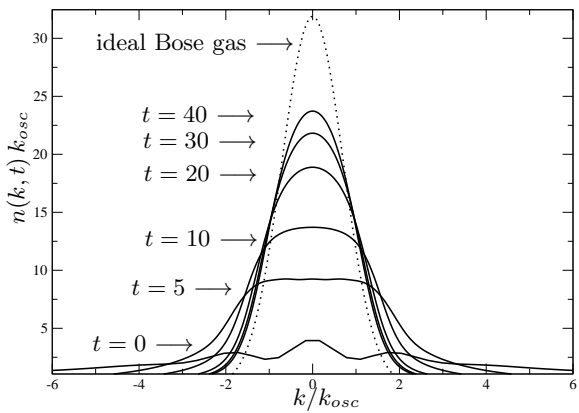} 
  \caption{Momentum distributions of a FTG gas (solid lines) with $N=9$ particles  as
  functions of the wavevector $k$  at subsequent times $t$ (in units
  of $1/\omega_0$) during a 1D expansion, and asymptotic long-time 
  expression (\ref{eq:bos}) (dashed line). 
}
  \label{fig2}
\end{figure}

{\it Expansion from a longitudinal harmonic trap:} We focus finally on 
a 1D expansion, as could be achieved by keeping on the transverse confinement.
If the 1D interactions are suddenly turned off before the gas is let free
to expand from a longitudinal harmonic trap, the density profile at long times 
reflects the initial momentum distribution \cite{GirWri05}. If instead the
interactions are kept on during the expansion we find that the density
profile expands self-similarly, while the momentum distribution
evolves from an initial overall Lorentzian shape \cite{BenErkGra04} to
that of an ideal Bose gas.
These properties can be demonstrated with the aid of an exact scaling
transformation as we outline below. Since the
FB mapping holds also for time-dependent phenomena induced by
one-body external fields \cite{time-dep}, the exact many-body wavefunction 
$\psi_F(x_{1},\cdots,x_{N};t)=A(x_{1},\cdots,x_{N})\prod_{j=1}^{N}
\phi_{0}(x_{j};t)$
during the dynamics is fully determined by the solution of the single-particle
Schr\"{o}dinger equation for the orbital $\phi_{0}(x_{j};t)$. For the
case of an external potential $V_{ext}(x,t)=m\omega(t)^2 x^2/2$ with
$\omega(0)=\omega_0$ the solution is known \cite{perelomov} to be
$\phi_0(x;t)=\phi_0(x/b(t);0)e^{imx^2\dot b/ 2b\hbar-iE_0\tau(t)/\hbar}$
where $b(t)$ is the solution of the differential equation $\ddot
b+\omega^2(t) b=\omega^2_0/b^3$ with $b(0)=1$ and $\dot b(0)=0$,
$\tau(t)=\int_0^t dt'\,1/b^2$ and $E_0=\hbar \omega_0/2$. Since the unit
antisymmetric wavefunction $A$ is invariant under the scaling
transformation, we immediately obtain the expression for the many-body
wavefunction,
$\psi_F(x_{1},..,x_{N};t)=b^{-N/2}\psi_F(x_{1}/b,..,x_{N}/b;0)
e^{i(\dot b/b\omega_0) \sum_{j=1}^N x_j^2/2x_{osc}^2}  
e^{-iNE_0 \tau(t)/\hbar}$,
and for the one-body density matrix,
$\rho_1(x,x';t)=\frac{1}{b}\rho_1\left(\frac{x}{b},\frac{x}{b};0\right) \exp\left[-i \frac{\dot b}{b}\frac{(x^2-x'^2)}{2x_{osc}^2}\right].$   
This yields the momentum distribution as a function
of time. While the intermediate-time dynamics has to be determined 
numerically, the stationary-phase method determines the
long-time evolution of the momentum distribution 
in the same way as for the bosonic TG gas \cite{MinGan05}. For the case of a 1D
expansion the scaling parameter is $b(t)=\sqrt{1+\omega_0^2t^2}$ and the
momentum distribution tends to that of an ideal Bose gas under
harmonic confinement, 
\begin{equation}
\label{eq:bos}
n(k,t\rightarrow\infty)\simeq  |\omega_0/\dot b| n_B(k\omega_0/\dot b),
\end{equation}
where $n_B(k)=2\pi N |\tilde \phi_0(k)|^2$, with $\tilde \phi_0(k)=\pi^{-1/4}k_{osc}^{-1/2}e^{-k^{2}/2k_{osc}^2}$ and $k_{osc}=1/x_{osc}$.
This behavior is illustrated in Fig.~\ref{fig2}. Quite noticeably, the
``bosonization'' time appears to be much longer than the ``fermionization''
time of the momentum distribution of the bosonic TG gas \cite{MinGan05}. 
Note that the ``dynamical bosonization'' described above
does not violate the Pauli exclusion principle: by using the above scaling solution
for the one-body density matrix  and  fixing unit
normalization of the natural orbitals at all times it follows that the
eigenvalues $\alpha_j$ of $\rho_1(x,x';t)$ are invariant during the expansion
%\cite{Note3} 
and hence always satisfy the condition $\alpha_j\le 1$.

In conclusion, we have found that (a) the untrapped system exhibits 
superconductive ODLRO of the two-body density matrix $\rho_2$ 
associated with its maximal eigenvalue $N$ and pair eigenfunction
$\mathcal{C}\ \text{sgn}(x_{1}-x_{2})e^{-2n|x_{1}-x_{2}|}$; 
(b) on a ring it has a highly degenerate ground state for even atom number,
and it exhibits quantization of rotational Coriolis flux 
and high sensitivity to rotation and to 
accelerations, making it a good candidate for high-sensitivity detectors;
(c) the harmonically trapped system undergoes a ``dynamical bosonization'' 
of its momentum distribution during a 1D expansion. 

\begin{acknowledgments}
This work was initiated at the Aspen Center for Physics during the
summer 2005 workshop ``Ultracold Trapped Atomic Gases''. We are grateful to the
organizers, G~Baym, R.~Hulet,
E.~Mueller, and F.~Zhou, for the opportunity to participate, and to
S.~Giorgini, R.~Seiringer, F.~Zhou,
E.~Zaremba, and G.~Shlyapnikov for helpful comments. 
The Aspen Center for Physics is supported by the
U.S. National Science Foundation, research of M.D.G. at the University of
Arizona by U.S. Office of Naval Research grant N00014-03-1-0427 through a 
subcontract from the University of Southern California, and that of
A.M. by  the Centre National de la
Recherche Scientifique (CNRS). 
\end{acknowledgments}

\begin{thebibliography}{33}
\bibitem{commentbc} This is the most convenient choice to pass to the the 
thermodynamic limit 
of an infinitely long linear trap containing an FTG gas of uniform density.
\end{thebibliography}
\end{document}